\documentclass[11pt,a4paper,reqno]{amsart}

\usepackage{times} 
\usepackage{amsmath} 
\usepackage{amssymb}  
\usepackage{amsthm}
\usepackage{latexsym}
\usepackage{amsfonts,bbm}
\usepackage{xcolor}
\usepackage{mathtools}
\usepackage{enumerate}
\usepackage{cite}
\usepackage{tikz}
\usepackage{graphicx}
\usepackage{microtype}
\usepackage{placeins}
\usepackage{float}
\usepackage{url}
\usepackage[calcwidth=0.8\linewidth]{caption}

\newcommand{\R}{\mathbb R}
\newcommand{\calI}{\mathcal I}
\newcommand{\calA}{\mathcal A}

\newcommand{\im}{\operatorname{im}}

\newtheorem{thm}{Theorem}[section]

\newtheorem{proposition}[thm]{Proposition}
\theoremstyle{definition}

\numberwithin{equation}{section}
\allowdisplaybreaks

\title{Energy-optimal control of adaptive structures}

\author[M.\ Schaller, A.\ Zeller, M.\ Böhm, O.\ Sawodny, C.\ Tarín and K.\ Worthmann]{Manuel Schaller$^{1}$,  Amelie Zeller$^{2}$, Michael Böhm$^{2}$, Oliver Sawodny$^{2}$, Cristina Tarín$^{2}$ and Karl Worthmann$^{1}$}
\thanks{}
\thanks{$^{1}$Optimization-based Control Group, Institute for Mathematics, Technische Universit\"at Ilmemau, Germany
	{\tt\small \{manuel.schaller,karl.worthmann\}@tu-ilmenau.de}.} %
\thanks{$^{2}$Institute for System Dynamics, University of Stuttgart, Germany
	{\tt \small  \{amelie.zeller,michael.boehm, oliver.sawodny,cristina.tarin-sauer\}@isys.uni-stuttgart.de}\\
$^*$ The authors gratefully acknowledge funding by the German Research Foundation (DFG) under the project number 279064222 (AZ, MB, OS and CT; SFB 1244, project B02) and under the project number 507037103 (KW). }%

\begin{document}
\begin{abstract}
Adaptive structures are equipped with sensors and actuators to actively counteract external loads such as wind. This can significantly reduce resource consumption and emissions during the life cycle compared to conventional structures. A common approach for active damping is to derive a port-Hamiltonian model and to employ linear-quadratic control. 
However, the quadratic control penalization lacks physical interpretation and merely serves as a regularization term. Rather, we propose a controller, which achieves the goal of vibration damping while acting energy-optimal. Leveraging the port-Hamiltonian structure, we show that the optimal control is uniquely determined, even on singular arcs. Further, we prove a stable long-time behavior of optimal trajectories by means of a turnpike property. Last, the proposed controller's efficiency is evaluated in a numerical study.

\end{abstract}

\vspace*{-.5cm}
\maketitle

\section{Introduction}
\noindent For the construction of stiffness-governed structures such as high-rise buildings, most of the material is required to provide the stiffness needed to meet the serviceability criteria for maximum displacement and acceleration under very extreme and hence rare external loads such as strong  \\[.06cm]
\begin{minipage}{.62\linewidth}
	winds or earthquakes. Without these extreme events, previous work on life-cycle analysis (i.\,e., including material and energy consumption and emissions during the construction, operation, and disposal) suggests potential resource savings of about~$50\%$~\cite{Schlegl2019}.
	To realize this enormous saving potential, the authors of~\cite{Teuffel01} suggest adaptive structures, which are equipped with sensors and actuators and can therefore meet the serviceability criteria by adaptation. These structures exhibit virtual stiffness through static compensation of displacements and virtual damping through actively controlled vibration mitigation. The world's first adaptive high-rise building was recently built at the campus of the University of Stuttgart, see Figure~\ref{fig:demonstrator}. It is 36\,m high and actuated by 24 hydraulic cylinders.  The operation of an adaptive structure requires it to actively react to external loads. This can be realized using modern model-based control techniques. 
	For modeling of complex systems across various energy domains, e.\,g., mechanical structures with hydraulic actuators, the port-Hamiltonian approach is especially beneficial. 
\end{minipage}
\begin{minipage}{.42\linewidth}
\vspace{.2cm}
\centering
\includegraphics[width=.8\columnwidth]{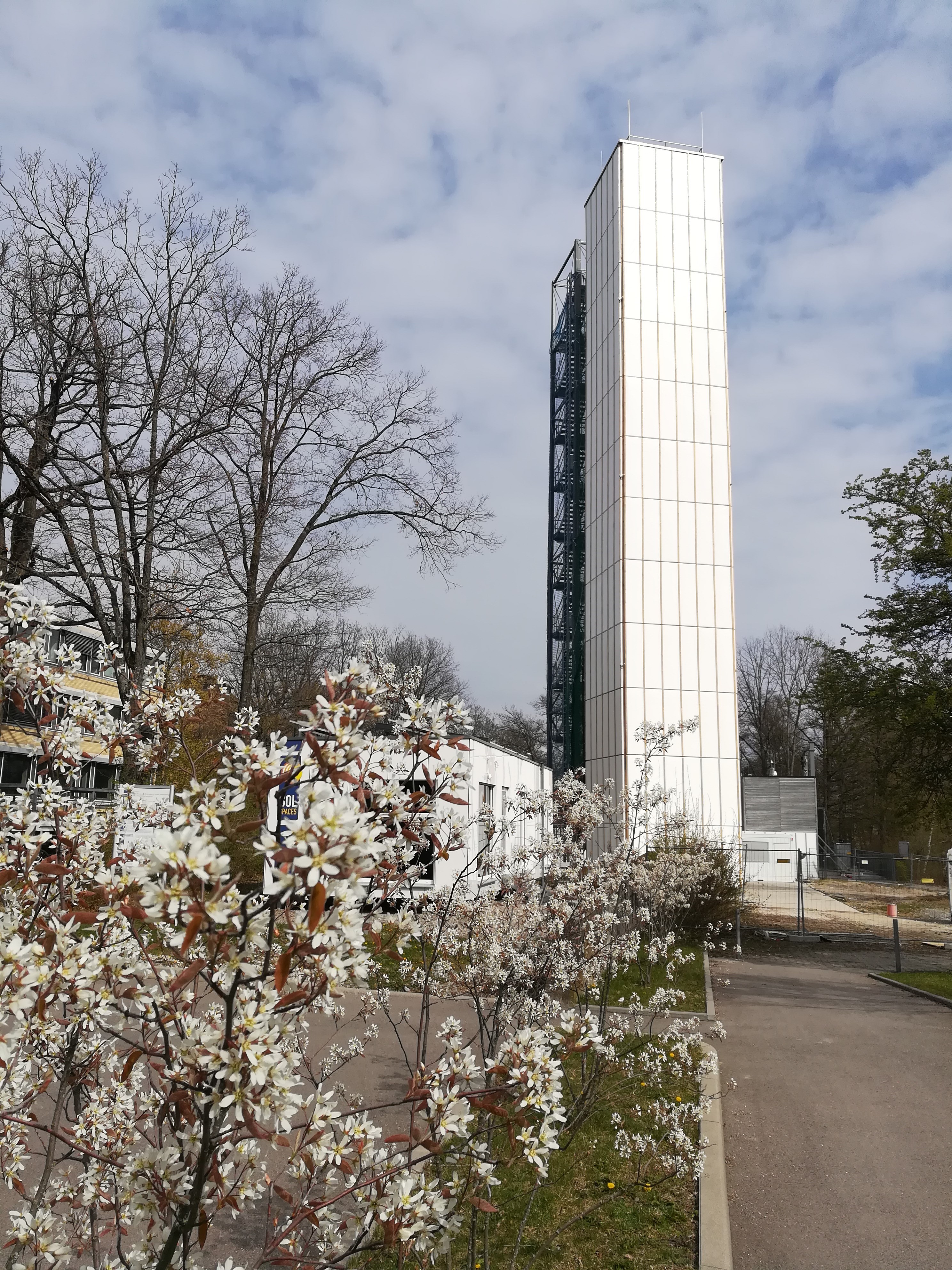}
\captionof{figure}{The world's first adaptive high-rise at the campus of the University of Stuttgart.}
\label{fig:demonstrator}
\end{minipage} \\[.06cm]
\newpage 
The reason for this is two-fold: First, the physical energy is directly taken into account in the mathematical model by means of the Hamiltonian. 
This allows for high interpretability of, e.g., analytical and numerical results. Second, the presence of ports, e.g., inputs and outputs, allows to couple subsystems and thus to model complex systems, such as adaptive high-rise buildings, in a modular way. Successful applications of the port-Hamiltonian structure range from functional analysis~\cite{JacoZwar12}, controller \cite{OrteScha02} and observer design~\cite{VenkScha10}, to (numerical) linear algebra \cite{MehlMehr18} or numerical discretization in time \cite{KotyLefe19} and space \cite{KotyMasc18}, to name just a few. 
%
%
Regarding adaptive high-rise structures, previous work considered port-Hamiltonian modeling and spatial discretization via (partitioned) finite elements~\cite{WarsBoeh21,Wars21}, as well as linear-quadratic controller design~\cite{WarsWagn20, WarsWagn20b, DakoHeid22}, where 
a weighted sum of the squared system state and the input integrated over time is minimized. 




In this paper, we suggest an optimization-based controller for adaptive structures which takes the port-Hamiltonian structure into account 
by maximizing the withdrawn energy of the system. 
Whereas this choice 
leads to an efficient controller in view of future implementations of energy harvesting, it also reduces natural dissipation.
As this natural dissipation takes place at large deformations and velocities and hence accelerates structural degradation, the suggested energy-optimal control is also favorable in view of durability. 

The optimal control problem of minimal energy supply (or equivalently maximal energy extraction) was introduced and analyzed in~\cite{SchaPhil21} for state-transition problems.
Due to the collocated structure of inputs and outputs in port-Hamiltonian modeling, the supplied energy is given as a time integral of the product of input and output. 
For adaptive structures, the inputs are the forces introduced by the hydraulic actuators, and outputs are formed by the corresponding velocities.  From a mathematical perspective this cost functional depends linearly on the control, such that the resulting optimal control problem is singular. For such singular control problems, standard results such as Riccati theory are not directly applicable and the optimality system can be singular in the control variable. This is due to the absence of a definite quadratic control penalization as required in standard Riccati theory, cf.\ \cite{Will71}. The port-Hamiltonian structure and the strong connection of cost and dynamics have been exploited in \cite{FaulMasc22,SchaPhil21} to show regularity of the pencil corresponding to the optimality system and stability of states and adjoints in optimal control of port-Hamiltonian ODEs and DAEs. In~\cite{PhilScha21} this was extended to infinite-dimensional systems including beam equations. 
Recently, for linear finite-dimensional ODEs, the regularity of the optimality pencil has been 
characterized in~\cite{FaulKirch23} with a corresponding state-feedback form of the optimal control.

The contribution of this work is twofold. 
First, to complement previous works considering state transitions, we provide a theoretical foundation for stabilizing optimal control with minimal energy supply.  
Resolving the singularity of the OCP by exploiting the port-Hamiltonian structure, we derive an analytical formula for the optimal control. Hereby, the control on singular arcs is fully determined by solving the respective optimality system making the dependence on state and adjoint explicit.  
Moreover, we show that, besides being energy optimal by construction, the corresponding optimal controls are stabilizing, analogously to their counterparts in case of quadratic control penalization. 
More precisely, we prove a turnpike property, showing that states corresponding to the energy-optimal controls are small in an integral sense.
In the second part of this work, we conduct 
a numerical case study, where we compare the energy-optimal controls with controls obtained from linear-
quadratic regulation.
In line with the theoretical results, this comparison is drawn by means of withdrawn energy and stability properties of the state.

This paper is organized as follows: In Section~\ref{sec:prelim}, we recall linear finite-dimensional port-Hamiltonian systems arising in modeling of adaptive buildings. In particular, we consider the respective power and energy balance, which is central for formulating the energy-optimal control problem. 
In Section~\ref{sec:theoretic}, we first prove a stable long-term behavior of of every optimal solution of the optimal linear-quadratic control problem 
by means of an exponential, i.e., pointwise, and an integral turnpike property. 
Then, we formulate an energy-optimal control problem which includes, instead of a quadratic control term, the supplied energy in the cost function. 
We derive an explicit formula of the optimal control using the optimality system and a stable long-term behavior in the sense of an integral turnpike property. 
In Section~\ref{sec:buildingcontrol}, we provide a numerical study by comparing two different cost functionals: First, we consider the penalization of the Hamiltonian and, second, we penalize the displacements of the upmost level. 
Last, in Section~\ref{sec:conclusion}, we summarize the main findings and conclude the paper by an outlook towards future work.

\section{Preliminaries} \label{sec:prelim}
In this part we introduce linear finite-dimensional port-Hamiltonian systems and recall how to obtain port-Hamiltonian models for adaptive structures.
\subsection{Port-Hamiltonian systems}
Linear port-Hamiltonian (pH) input-state-output systems are defined as
\begin{align}\tag{pH}\label{eq:pH}
\dot{x}(t) = (J-R)Qx(t) + Bu(t),\quad y(t) = B^\top Qx(t).
\end{align}
Here, $J=-J^\top\in \mathbb{R}^{n\times n}$ is a structure matrix corresponding to energy-routing in the system, the matrix $0\leq R=R^\top \in \mathbb{R}^{n\times n}$ encodes the internal energy dissipation and $B\in \mathbb{R}^{n\times m}$ is an input matrix. The matrix $0\leq Q=Q^\top\in \mathbb{R}^{n\times n}$ captures the energy storage by means of the Hamiltonian $H(x) := \frac{1}{2} x^\top Q x$.  

Due to the structure of the involved matrices, the trajectories of \eqref{eq:pH} satisfy a power balance. More precisely, for all $t\geq 0$
\begin{align}\label{eq:energy_balance}
\underbrace{\tfrac{\mathrm{d}}{\mathrm{d}t}H(x(t))}_{\text{power flow to storage}} \!\!\!\!=\! -\underbrace{\|R^{1/2}Qx(t)\|^2}_{\text{dissipated power}} + \underbrace{u(t)^\top y(t)}_{\text{supplied power}}.
\end{align}
If we integrate this equation, we obtain the corresponding energy balance, i.e., for all $t\geq 0$,
\begin{align*}
\underbrace{H(x(t)) - H(x(0))}_{\text{change of energy}}= -\underbrace{\int_0^t \|R^{1/2}Qx(s)\|^2\,\mathrm{d}s}_{\text{dissipated energy}} + \underbrace{\int_0^t u(s)^\top y(s)\,\mathrm{d}s}_{\text{supplied energy}}.
\end{align*}
From these equations it can be seen that the supplied energy to the system is given by the \textit{passivity supply rate} $u(t)^\top y(t)$. Regardless of the physical domain, e.g., mechanics, electrical engineering or fluid dynamics, cf.\ \cite{SchaJelt2014}, the product of input and output will always have the physical unit of power, allowing for a coupling of port-Hamiltonian systems across different domains. If this coupling is performed in a power-preserving manner, the coupled system can be shown to be again port-Hamiltonian.

\subsection{Port-Hamiltonian modeling of adaptive buildings}
This section describes the modeling of adaptive structures, such as adaptive high-rise buildings. Structures comprising beams and links can be seen as flexible multi-body systems with infinitely many degrees of freedom. Such models include, e.g., Timoshenko or Euler-Bernoulli beam models and are given in the form of a partial differential equation.
To obtain a model with finitely many degrees of freedom~(DOFs), the finite element method is commonly applied. In this regard, a port-Hamiltonian model can be obtained by two means.
First, the structural elements (links and beams here) are each modeled in port-Hamiltonian form, then spatially discretized using the partitioned finite element method (PFEM) from \cite{Cardoso2021} and last assembled to a complex structure \cite{WarsBoeh21}. 
Second, a model in port-Hamiltonian form can be obtained by first deriving the linear second order equations of motion of the whole structure using the conventional finite element method (FEM) and then transforming it to port-Hamiltonian form.


Using the conventional FEM, the structures are discretized by finite beam or link elements, which are connected at nodes.
Each node has a certain number of translational and rotational DOFs, depending on the configuration and used types of finite elements (e.g., links need only translational DOFs), the dimensionality (e.g.,two or three-dimensional) and constraints (e.g., at the support and coupling interfaces).
The DOFs are collected in the 
vector $q(t) \in\mathbb{R}^{n_\mathrm{dof}}$ and the linear second order equations of motion are, for given initial DOFs $q_0\in \R^{n_\mathrm{dof}}$ and velocities $v_0\in \R^{n_\mathrm{dof}}$, given by
\begin{subequations}
	\label{eq:femsys}
	\begin{align}
	&M\ddot{q}(t)  + D\dot{q}(t) + K q(t) = f(t),\ t>0, \\
	&q(0) = q_0,\ \dot{q}(0) = v_{0}
	\end{align}
\end{subequations}
where $\dot{q}(t)$ are the nodal velocities, $\ddot{q}(t)$ are nodal accelerations and $M, K\in\mathbb{R}^{n_\mathrm{dof}\times n_\mathrm{dof}}$ are the positive definite mass and stiffness matrices, respectively.

Any kind of civil engineering structure exhibits some kind of damping.
Rayleigh-damping as introduced in~\cite{Strutt1877} is a common damping model in the literature.
This is due to its numerically beneficial assumption that damping is a linear combination of mass and stiffness, $D = \alpha_1 M + \alpha_2 K$. For this choice, the undamped and the damped system exhibit identical eigenmodes. For the simulations in this paper we use $\alpha_1 = 0.05, \alpha_2= 0.005$. For the proposed control method, a different damping model satisfying $D=D^\top\succ 0$ would also be feasible.
The forces acting on the structure are collected in $f(t)\in\mathbb{R}^{n_\mathrm{dof}}$ and comprise external loads, such as wind loads and the forces applied by the actuators.
In this work, no external loads are considered and a structure is excited by a non-zero initial condition, corresponding, e.g., to a previous displacement by wind. 

The actuators are integrated in parallel, meaning that the actuator forces act in parallel to the hosting elements in opposite directions at the adjacent nodes. Thus $f(t)$ is rewritten as $f(t) = F_{\mathrm{u}}u(t)$, where $u(t)\in\mathbb{R}^m$ contains the actuator forces (the input) and the input matrix $F_{\mathrm{u}}\in\mathbb{R}^{{n_\mathrm{dof}}\times m}$ maps the input to nodal forces.
For simplicity of notation, the time dependency and the initial conditions are omitted in the
following.



The model equations \eqref{eq:femsys} are transformed to port-Hamiltonian form \eqref{eq:pH} using the DOFs and momenta as state variables $x= \left(\begin{smallmatrix}M\dot{q}\\q\end{smallmatrix}\right) \in \mathbb{R}^{2n_\mathrm{dof}}$:
\begin{align}\label{eq:sosys}
\dot{x} \!= \!\left( \underbrace{\begin{bmatrix}
	0&-I\\I&0
	\end{bmatrix}}_{J}
-\underbrace{\begin{bmatrix}
	D&0\\
	0&0
	\end{bmatrix}}_{R}\right)
\underbrace{\begin{bmatrix}
	M^{-1}&0\\
	0&K
	\end{bmatrix}}_{Q}
x
+ 
\underbrace{\begin{bmatrix}
	F_{\mathrm{u}}\\0
	\end{bmatrix}}_{B} u.
\end{align}
Thus, the energy 
\begin{align*}
H(x) = \frac12\begin{pmatrix} (M\dot {q})^\top \, q^\top\end{pmatrix} \begin{bmatrix}
M^{-1}&0\\
0&K
\end{bmatrix} \begin{pmatrix}
M\dot q\\ q
\end{pmatrix} = \frac12\left(\dot q^\top M \dot q + q^\top K q\right)
\end{align*}
consists of kinetic and potential energy. Further, the collocated output according to \eqref{eq:pH} is defined by
\begin{align}
y = \underbrace{\begin{bmatrix} F_{\mathrm{u}}^\top & 0 \end{bmatrix}}_{B^\top} Q x = F_{\mathrm{u}}^\top \dot{q}.
\end{align}
and has the physical unit of a velocity.



\section{Optimal control of adaptive buildings} \label{sec:theoretic}


In this work, we will consider two different OCPs in view of damping performance and energy efficiency. 
The cost functionals of both OCPs will include a term $x^\top W x$, $W\in \mathbb{R}^n$ with $W=W^\top \geq 0$ to penalize the state variables. 
In particular we have the following choices in mind:
\begin{itemize}
	\item[1)] First, we set $W = Q$, that is, we penalize the Hamiltonian over time via $x^\top W x = 2H(x)$.
	\item[2)] Second, we choose $W$ such that it penalizes the Hamiltonian corresponding to the upmost level.
\end{itemize}
Due to invertibility of $D$ in \eqref{eq:sosys}, i.e., damping is present in all components, $A = (J-R)Q$ is a Hurwitz matrix. Thus any output matrix gives rise to a detectable output (by choosing zero as a feedback matrix), i.e., in particular, $(A,W^{1/2})$ is detectable. Analogously, any input matrix~$B$ renders $(A,B)$ stabilizable. In view of the control cost, we will first consider quadratic penalizations given by $\mu \int_0^T \|u(t)\|^2\,\mathrm{d}t$, $\mu > 0$ and penalizations of the supplied energy given by $\int_0^T u(t)^\top y(t)\,\mathrm{d}t$ in view of \eqref{eq:energy_balance}.

\subsection{Linear quadratic optimal control} 

First, we inspect solutions of a linear quadratic problem given by
\begin{subequations}\label{eq:lqrocp}
	\begin{align}
	\min_{u\in L^1(0,T;\mathbb{U})} &\int_0^T x^\top W x + \mu \|u(t)\|^2\,\mathrm{d}t\\
	\text{s.t. } &\eqref{eq:pH},\quad x(0)=x^0.
	\end{align}
\end{subequations}
where $x^0 \in \mathbb{R}^{n}$ is an initial state, $\mu > 0$ is a regularization parameter and $\mathbb{U}\subset \R^m$ is a compact and convex control constraint set. In particular, $\mathbb{U}$ is bounded such that $L^p(0,T;\mathbb{U})\subset L^\infty(0,T;\mathbb{U})$ for all $1\leq p\leq \infty$. 

In the remainder of this section, we state two theoretical results w.r.t.\ the long-time behavior of \eqref{eq:lqrocp} in the form of a turnpike property. Such turnpike properties are crucial in the analysis of model predictive control schemes \cite{Grun13,GrunPann17}.
We first consider the case of inactive control constraints, which, in view of Riccati theory is present if the penalization parameter $\mu > 0$ is large w.r.t.\ the norm of the initial value.
\begin{proposition}\label{prop:regular_integraltp}
	Let $0\in \mathbb{U}$. 
	If the control constraints in~\eqref{eq:lqrocp} are inactive at all times, then the OCP~\eqref{eq:lqrocp} has an exponential turnpike property towards zero, i.e., there is a decay rate $\omega> 0$ and a constant $c\geq 0$ such that for all initial values $x^0\in \mathbb{R}^n$ and time horizons $T\geq 0$, the optimal state control pair $(x^*,u^*)\in W^{1,1}(0,T;\mathbb{R}^n)\times L^1(0,T;\mathbb{R}^m)$ satisfies
	\begin{align*}
	\|x^*(t)\|^2 + \|u^*(t)\|^2 \leq c(e^{-\omega t} + e^{-\omega (T-t)})\|x^0\|.
	\end{align*}
	for almost every $t\in[0,T]$.
\end{proposition}
\begin{proof}
	As the control constraints are inactive, we can omit them in the optimal control problem \eqref{eq:lqrocp} and thus optimize over inputs $u\in L^1(0,T;\R^m)$.
	Correspondingly, the static optimal control problem with steady state pair $(\bar{x}^*,\bar{u}^*) \in \mathbb{R}^{n} \times \mathbb{R}^{m}$ is given as solution of
	\begin{align*}
	\min_{\bar{u}\in \R^m} \bar{x}^\top W \bar{x} + \mu\|\bar{u}\|^2, \quad \text{s.t.  } 0 =A\bar{x}+B\bar{u}
	\end{align*}
	where $A=(J-R)Q$. By standard KKT-conditions, there is an associated Lagrange multiplier $\bar{\lambda}\in \R^n$ such that the optimal pair ($\bar{x}^*,\bar{u}^*)\in \R^n\times \R^m$ satisfies
	\begin{align}\label{eq:staticoptsys}
	\begin{pmatrix}
	W & 0& A^\top\\
	0&\mu I& B^\top\\
	A&B&0
	\end{pmatrix}
	\begin{pmatrix}
	\bar{x}^*\\\bar{u}^*\\\bar{\lambda}
	\end{pmatrix}
	= \begin{pmatrix}
	0\\0\\0
	\end{pmatrix}.
	\end{align}
	As $A= \left(\begin{smallmatrix}
	-D& -I\\
	I&0
	\end{smallmatrix}\right) \left(\begin{smallmatrix}
	M^{-1}&0\\
	0&K
	\end{smallmatrix} \right)$ is invertible, the unique optimal steady state control $\bar{u}^*$ is the solution of the strictly convex problem 
	\begin{align*}
	\min_{\bar{u}\in \R^m} (A^{-1}B\bar{u})^\top W A^{-1}B\bar{u} + \mu \|\bar{u}\|^2
	\end{align*}
	and thus satisfies $\bar{u}^*=0$ with corresponding state $\bar{x}^* = 0$. By means of the first row of the optimality system~\eqref{eq:staticoptsys} and from invertibility of $A$, also the steady state adjoint vanishes.
	As $A$ is Hurwitz, $(A,B)$ is trivially stabilizable and $(A,W^{1/2})$ is detectable so that we can invoke \cite[Theorem 3]{TrelZhan18} to obtain a turnpike property with respect to the optimal steady state $(\bar{x},\bar{u})$
	\begin{align*}
	\|x^*(t)-\bar{x}^*\|^2 + \|u^*(t)-\bar{u}^*\|^2 \leq c(e^{-\omega t} + e^{-\omega (T-t)})\left(\|x^0-\bar{x}^*\| + \|\bar{\lambda}\|\right).
	\end{align*}
	Inserting the previous computation $(\bar{x}^*,\bar{u}^*) = 0$ shows the claim.
\end{proof}
Complementary to the case of inactive control constraints considered in Proposition~\ref{prop:regular_integraltp}, we now provide a result which also applies to the case of active control constraints.
\begin{proposition}\label{prop:intturnpike_LQR}
	Suppose that $0 \in \operatorname{int} \mathbb{U}$. Furthermore, let $W \in \R^{n\times n}$ be invertible and the pair $((J-R)Q,B)$ be controllable in the sense of Kalman. Then, the OCP~\eqref{eq:lqrocp} has an integral turnpike property, i.e., for all $x^0 \in \mathbb{R}^n$ there is a constant $c \geq 0$ such that for all $T> 0$, the optimal state control pair $(x^*,u^*)\in W^{1,1}(0,T;\mathbb{R}^n)\times L^1(0,T;\mathbb{R}^m)$ satisfies
	\begin{align}\label{eq:intturnpike}
	\int_0^T \|x^*(t)\|^2 + \|u^*(t)\|^2\,\mathrm{d}t \leq c.
	\end{align}
\end{proposition}
\begin{proof}
	By controllability, and as $0\in \operatorname{int}\mathbb{U}$, there is $u \in L^1(0,T;\mathbb{U})$ and $t_1\geq 0$ depending on $x^0$ such that $x(t;x^0,u)=0$ and $u(t)=0$ for all $t\geq t_1$ \cite{FaulMasc22}. Here, $x(t;x^0,u)$ denotes the state emanating from the dynamics \eqref{eq:pH} and the initial state $x^0$ when applying the control~$u$. 
	This control is constructed using reachability of the zero state, which follows from the spectral properties of the matrix $(J-R)Q$ and the fact that $0\in \operatorname{int}\mathbb{U}$, cf.\ \cite{FaulMasc22}. 
	W.\,l.\,o.\,g., we may assume that $T\geq t_1$. Then, we can conclude by optimality that the optimal state-control pair $(x^*,u^*)$ satisfies
	\begin{align}
	\int_0^T x^*(t)^\top Wx^*(t)  + \mu\|u^*(t)\|^2\,\mathrm{d}t
	&\leq \int_0^T \! x(t;x^0,u)^\top W x(t;x^0,u)+ \mu\|u(t)\|^2\,\mathrm{d}t \nonumber\\
	&=\int_0^{t_1} \! x(t;x^0,u)^\top W x(t;x^0,u) + \mu \|u(t)\|^2\,\mathrm{d}t\nonumber\\& \leq \tilde{c}(t_1),\label{eq:formula3}
	\end{align}
	where $\tilde{c}(t_1)\geq 0$ is a constant independent of $T$. Using invertibility of $W$ we have that $\lambda_\mathrm{min}(W) \|x\|^2 \leq x^\top W x$ 
	for all $x\in \mathbb{R}^n$, where $\lambda_\mathrm{min}(W)>0$ denotes the smallest eigenvalue of the invertible matrix $W$. Thus, we have
	\begin{align*}
	\int_0^T x^*(t)^\top Wx^*(t)  + \mu\|u^*(t)\|^2\,\mathrm{d}t \geq \min\{\lambda_{\mathrm{min}}(W),\mu\} \int_0^T \|x^*(t)\|^2 + \|u^*(t)\|^2\,\mathrm{d}t.
	\end{align*}
	\noindent  Now, inserting this into \eqref{eq:formula3} and dividing by $\min\{\lambda_{\mathrm{min}}(W),\mu\} > 0$ yields \eqref{eq:intturnpike} with $c = \frac{\tilde c(t_1)}{\min\{\lambda_{\mathrm{min}}(W),\mu\}}$.
\end{proof}
We briefly provide an interpretation of the integral turnpike property of the previous proposition, i.e., Inequality~\eqref{eq:intturnpike}.
As the upper bound is independent of $T$ and as the integral is nonnegative, this means that if we increase the time horizon and thus the integration domain, the integrand on the left hand side has to be small for the majority of the time. This means, that such a bound provides us with a notion of smallness of the optimal state and control. Moreover, passing to the limit via $T\to \infty$ in~\eqref{eq:intturnpike} we obtain $x\in L^2(0,\infty;\R^n)$ which implies the convergence $x(t) \stackrel{t\to\infty}{\to} 0$, cf.\ \cite[Lemma 14.1]{ZhouDoyl96}.

Having provided stability results for the quadratic control penalization, we now introduce an alternative cost functional, which takes the port-Hamiltonian structure into account.
\subsection{Optimal control with minimal energy supply}

In order to include the pH-structure in the cost function and to minimize the supplied energy, cf.\ \eqref{eq:energy_balance} we replace the quadratic control term in \eqref{eq:lqrocp} by the supplied energy. Thus, for $x^0 \in \mathbb{R}^{n}$, we consider the problem
\begin{subequations}\label{eq:phocp}
	\begin{align}
	\min_{u\in L^1(0,T;\mathbb{U})} &\int_0^T x(t)^\top W x(t) +  u(t)^\top y(t) \,\mathrm{d}t\\
	\text{s.t. } &\eqref{eq:pH},\quad x(0)=x^0.
	\end{align}
\end{subequations}
For the case of energy-optimal state transitions, that is, in case $W=0$ and an additional terminal condition $x(T)=x_T\in \mathbb{R}^n$, this problem was analyzed in \cite{FaulMasc22,SchaPhil21,FaulKirch23}.

The integrand of the cost function of problem \eqref{eq:phocp} is a particular case of the cost function of the LQR problem with cross-weighted costs as described in \cite{Levine2011} with a certain choice of weighting matrices. In view of the the notation in \cite{Levine2011}, the cost function for such cross-weighted stage costs is given by $x(t)^\top W x(t) + 2 x(t)^\top W^{1/2} D u(t) + u(t)^\top (P + D^\top D) u(t)$. The cost function integrand in \eqref{eq:phocp} results using $D = \frac{1}{2} W^{-1/2}QB$ and $P = -D^\top D$. Here it is important to note that we do not have a quadratic control penalization, which leads, in general, to a singular optimal control problem. Invoking the port-Hamiltonian structure allows to resolve this issue and fully characterize the optimal state feedback, see~\cite{FaulKirch23}.

Using the balance equation~\eqref{eq:energy_balance}, the cost of~\eqref{eq:phocp} can be cast into the form
\begin{align*}
&\int_0^T x(t)^\top W x(t) +  u(t)^\top y(t) \,\mathrm{d}t
= H(x(T)) -H(x(0)) \\
&\qquad\qquad +\int_0^T x(t)^\top W x(t) + \|R^{1/2}Qx(t)\|^2 \,\mathrm{d}t.
\end{align*}
As $H(x(0)) = H(x^0)$, we will omit this term in the following, as it poses a mere offset of the cost functional.
By means of the above reformulation, we see that the cost functional in \eqref{eq:phocp} implicitly carries a terminal cost by means of the Hamiltonian. Moreover, a particular focus on the velocity components is intrinsically included by means of the dissipation matrix $RQ$ of \eqref{eq:sosys}.

Note that in \eqref{eq:phocp} we did not include a penalization parameter $\mu > 0$ for the control cost as for the quadratic cost in \eqref{eq:lqrocp}. The reason for this is simple: as can be observed in the numerical results in Section~\ref{sec:buildingcontrol}, we do not need to include a tuning of this cost functional to observe active damping. This is contrary to the quadratic control penalization in \eqref{eq:lqrocp} where for too large penalization parameters no vibration damping beyond the natural dissipation is obtained.

First, similar to Propositions~\ref{prop:regular_integraltp} and \ref{prop:intturnpike_LQR}, we provide a result showing that energy-optimal solutions, i.e., solutions of \eqref{eq:phocp} also stabilize the system. 
To this end, we denote the distance to a closed subset $M\subset \mathbb{R}^{n}$ by $\operatorname{dist}(x,M) = \min_{z\in M}\|x-z\|$.
\begin{proposition}
	Let $0\in \operatorname{int}\mathbb{U}$ and assume that $((J-R)Q,B)$ is controllable in the sense of Kalman. Then, optimal states $x^*\in W^{1,1}(0,T;\mathbb{R}^n)$ of \eqref{eq:phocp} exhibit a subspace integral turnpike property w.\,r.\,t.\ $\ker RQ\cap \ker W$, i.e., for any initial value $x^0\in \mathbb{R}^n$ there is a constant $c\geq 0$ independent of $T$ such that
	\begin{align*}
	\int_0^T \operatorname{dist}(x^*(t),\ker RQ \cap \ker W)^2\,\mathrm{d}t \leq c.
	\end{align*}
	In particular, if $W$ is invertible, then 
	\begin{align*}
	\int_0^T \|x^*(t)\|^2\,\mathrm{d}t\leq c.
	\end{align*}
\end{proposition}
\begin{proof}
	Analogously to the proof of Proposition~\ref{prop:intturnpike_LQR}, we can deduce the estimate 
	\begin{align}\label{eq:subspace_tp_proof}
	\int_0^{t_1} \|R^{1/2}Q x^*(t) \|^2 + (x^*(t))^\top W (x^*(t)) \leq c(t_1),
	\end{align}
	where $c(t_1)\geq 0$ is a constant independent of the time horizon $T$. Then there are constants $c_1,c_2,c_3\geq 0$ such that for all $x\in \R^n$
	\begin{align*}
	\|R^{1/2} x \|^2 + \|W^{1/2}x\|^2 &\geq 
	c_1\operatorname{dist}(x, \ker R^{1/2} Q)^2 \!+\!    c_2\operatorname{dist}(x,\ker W)^2\\
	&\geq c_3\operatorname{dist}(x,\ker R^{1/2} Q \cap \ker W)^2,
	\end{align*}
	where in the first inequality we used \cite[Lemma 13]{SchaPhil21} and in the second inequality we invoked \cite[Lemma A.1]{PhilScha23}.
	Inserting this into the estimate \eqref{eq:subspace_tp_proof}, the first claim follows. For the second claim, by invertibility of $W$ we get that $\ker W = \{0\}$ such that $\ker RQ \cap \ker W = \{0\}$ and hence, $\operatorname{dist}(x^*,\ker RQ \cap \ker W) = \operatorname{dist}(x^*,0) = \|x^*\|$.
\end{proof}

For the quadratic control cost in \eqref{eq:lqrocp}, by standard Riccati theory and in the absence of control constraints, the control is given as a state feedback. For finite time horizons, this feedback can be obtained solving the differential Riccati equation, and on the infinite time horizon, by the algebraic Riccati equation, respectively. Due to the linear dependence of the cost on the control, such results can not be applied to \eqref{eq:phocp}. In the following, however, we show under a condition on the input matrix $B$, the dissipation matrix $RQ$ and the state weighting matrix $W$ in the cost that the control is given by means of state and adjoint state. This paves the way for future work considering state-feedback controls, as provided in \cite{FaulKirch23} for state transition problems.

To derive optimality conditions, we define the optimality Hamiltonian
\begin{align*}
\mathcal{H} &: \R^n\times\R^m\times\R_{\ge 0}\times\R^n \to \mathbb{R}\\
(x,u,\lambda_0,\lambda) &\mapsto \lambda_0\!\cdot\! \left(x^\top W x + u^\top \!y\right) \!+ \!\lambda^\top\! \left((J-R)Qx\!+\!Bu\right).
\end{align*}
By Pontryagin's maximum principle, cf., e.g., \cite{Lee67}, optimal solutions $(x^\star ,u^\star )\in W^{1,1}(0,T;\mathbb{R}^n)\times L^1(0,T;\mathbb{U})$ of~\eqref{eq:phocp} satisfy 
\begin{subequations}
	\label{eq:OC}
	\begin{align}
	\dot{x}^\star (t) &= \phantom{-}\mathcal{H}_\lambda(x^\star (t),u^\star (t),\lambda_0,\lambda(t))\\
	\dot{\lambda}(t) &= -\mathcal{H}_x(x^\star (t),u^\star (t),\lambda_0,\lambda(t))\\
	u^\star (t)& \in \arg\min_{\tilde{u}\in \mathbb{U}}  \phantom{-}\mathcal{H}(x^\star (t),\tilde{u},\lambda_0,\lambda(t)). \label{eq:OC_minH}
	\end{align}
\end{subequations}
In view of \cite[Rem. 6.9, p. 168]{Locatelli01}, as we do not impose any terminal constraints, the multiplier $\lambda_0$ does not vanish such that we may normalize $\lambda_0 =1$ in the following. Inserting the output equation $y = B^\top Q x$
\begin{align*}
\mathcal{H}(x,u,\lambda)= \,&u^\top B^\top (Qx +\lambda) + x^\top W x+ \lambda^\top (J-R)Qx
\end{align*}
and thus we obtain the optimality system 
\begin{align}
\dot{x}^*(t) &= \phantom{-}(J-R)Qx^*(t) + Bu^*(t)\nonumber \\
\dot{\lambda}(t)&= -QBu^*(t) - 2Wx^*+ Q(J+R)\lambda(t)\label{eq:OC_2}\\
u^*(t) &\in \arg\min_{\tilde{u}\in \mathbb{U}}\,\tilde u^\top B^\top (Qx^*(t)+\lambda(t)).\nonumber
\end{align}
For components of the switching function
$$
s(t) = B^\top(Qx^*(t)+\lambda(t))
$$
which do not vanish, the corresponding optimal control components are determined by means of a linear minimization problem and lie on the boundary of the compact control constraint set $\mathbb{U}$.
However, if the $i$-th component of the switching function 
vanishes, the inclusion in the last line of \eqref{eq:OC_2} does not provide any information on the control component $u_i$, $i\in \{1,\ldots,m\}$. Thus, time intervals of positive measure where $s_i(t) \equiv 0$, $i\in \{1,\ldots,m\}$ are called singular arcs.
On these singular arcs and on first sight, the optimality condition does not characterize the optimal control. The following result shows that under a certain condition linking the input matrix, the dissipation matrix, and the state weighting in the cost, this singularity can be resolved. More precisely, the optimal control on singular arcs can be expressed in terms of the state and the adjoint.
\begin{thm}\label{prop:u_xla}
	Assume $\im B\cap\ker RQ\cap \ker W = \{0\}$ and that $(x^*,u^*,\lambda)$ satisfies the optimality system~\eqref{eq:OC_2} of the OCP~\eqref{eq:phocp}.
	Then the optimal control $u^*$ is completely determined by $x^*$ and $\lambda$. 
	
	\noindent More precisely, for a given interval~$I$, we consider a partition of the index set~$\{1,\ldots,m\}$ into switching indices
	$$
	\calI \doteq \{i : s_i \equiv 0 \text{ on $I$}\} \quad\text{and}\quad \calA \doteq \{i : s_i \not\equiv 0 \text{ a.e.\ on $I$}\}.
	$$
	Set $u_\calI = (u_i)_{i\in\calI}^\top$, $u_\calA = (u_i)_{i\in\calA}^\top$, $B_\calI = (b_i)_{i\in\calI}$, and $B_\calA = (b_i)_{i\in\calA}$. %
	Then, on a singular arc $I$, we have
	$$
	u_\calI = M^{-1}B_\calI^\top\big[\tfrac 12((QA^2-2W)x + (A^2)^\top\lambda) - QRQB_\calA u_\calA\big],
	$$
	where $M = B_\calI^\top QRQB_\calI$ and $A = (J-R)Q$.
\end{thm}
\begin{proof}
	Let $s_\calI = B_\calI^\top(Qx+\lambda)$. Then $s_\calI\equiv 0$ on $I$. Hence,
	$$
	0 = \dot s_\calI = B_\calI^\top(Q\dot x+\dot\lambda) = B_\calI^\top(QAx-A^\top\lambda - 2Wx).
	$$
	Taking the derivative w.r.t.\ time once again on $I$ and setting $v \doteq  \tfrac 12((QA^2 - 2WA-2A^\top W)x + (A^2)^\top\lambda)$ gives
	\begin{align*}
	0
	&= \ddot s_\calI = B_\calI^\top(QA\dot x-A^\top\dot\lambda-2W\dot{x})\\
	&= B_\calI^\top\big[(QA-2W)(Ax+Bu) - A^\top(-QBu-2Wx - A^\top\lambda)\big]\\
	&= B_\calI^\top\big[2v + (QA+ A^\top Q-2W)Bu\big]\\
	&= 2B_\calI^\top v - 2B_\calI^\top (QRQ+W)Bu,
	\end{align*}
	as $QA+ A^\top Q = -2QRQ$. Thus, we obtain
	\begin{align*}
	B_\calI^\top (QRQ+W)B_\calI u_\calI = B_\calI^\top v - B_\calI^\top (QRQ+W)B_\calA u_\calA.
	\end{align*}
	As $QRQ$ and $W$ are symmetric positive semidefinite, we have $\ker(QRQ+W)=\ker QRQ \cap \ker W = \ker RQ \cap \ker W$.
	Hence, by our assumption $\im B \cap \ker RQ \cap \ker W =\{0\}$, the matrix $B_\calI^\top (QRQ+W)B_\calI$ is positive definite and thus invertible, which proves the claim. 
\end{proof}

We conclude this section by a brief discussion of the assumption $\im B \cap \ker RQ \cap \ker W = \{0\}$ of Theorem~\ref{prop:u_xla}. This condition is always satisfied (i.e., in particular for $W=0$) for fully damped (i.e., $D\succ 0$) second order systems of the form \eqref{eq:sosys}, as here already $\im B \cap \ker RQ = \{0\}$. This is a consequence of direct actuation only in the damped coordinates, i.e., the momenta.
For $W=0$, it has been shown in \cite{FaulKirch23} that the condition $\im B \cap \ker RQ = \{0\}$ is only sufficient and not necessary for characterizing the control by the state and the adjoint. More precisely, assuming full column rank of $B$, $\im B \cap \ker RQ = \{0\}$ holds if and only if the optimality system along the singular arc is a regular differential algebraic equation (DAE) with index three \cite[Proposition 3.10]{FaulKirch23}. We note that formula for the optimal control can be explicitly stated if the resulting DAE is regular, without a condition on the index \cite{FaulKirch23}.

\section{Active vibration damping of the high-rise building} 
\label{sec:buildingcontrol}

In this part, we provide two numerical case studies for active vibration damping via optimal control of adaptive buildings. In this context, we aim to asymptotically stabilize the system \eqref{eq:sosys} from an initial displacement to the rest state by means of the optimal control problems \eqref{eq:lqrocp} and \eqref{eq:phocp}. Whereas such a stabilization (or vibration damping) by means of quadratic costs as in \eqref{eq:lqrocp} is standard, we will show that the formulation \eqref{eq:phocp} provides advantages in view of the withdrawn energy, which is explicitly taken into account in the cost functional. 

\begin{figure}[htb]
	\centering
	\includegraphics[width=0.35\columnwidth]{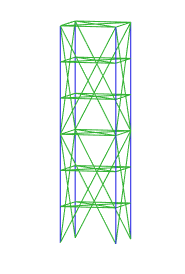}
	\caption{Slender high-rise type example structure with six storeys. Structural elements colored blue are beams, structural elements colored green are links.}
	\label{fig:structure}
\end{figure}

The case study will be performed by means of two choices of the state weighting matrix as formulated in Section~\ref{sec:theoretic}. First, we want to penalize the full Hamiltonian, second we aim to penalize Hamiltonian of the building's upmost storey. 
Exemplarily, we consider a slender six storey structure, see Figure \ref{fig:structure}, consisting of 28 nodes of which four are ground nodes.
\noindent The vertical columns are beams all other structural elements are links (colored blue and green in Figure~\ref{fig:structure}, respectively).
Each upper node has six DOFs, while the displacements and the rotation around the vertical axis of the ground nodes are constrained. Therefore, the ground nodes have only two DOFs. In total, the model has $4\cdot 2 + 24\cdot 6 = 152$ DOFs and equally many momenta, leading to a state dimension $2\cdot 152 = 304$.
A computational model of the form \eqref{eq:pH} is obtained by means of the Matlab framework introduced in \cite{WarsBoeh21} available on GitHub\footnote{\url{https://github.com/awarsewa/ph_fem/}}. 
Actuators are placed in the vertical columns spanning over the three storeys. In addition, all diagonals spanning over three storeys are actuated. All actuators are placed in parallel to their hosting elements. The maximal force of the actuators is set to $u_\mathrm{max}=100$\,kN as in \cite{DakoHeid22}, i.e., $\mathbb{U}=[-10^{5},10^{5}]^m$.

The resulting dynamics is discretized by means of an implicit midpoint rule as a symplectic integrator on a time grid with 301 time grid points on the horizon $T=1$. The corresponding optimization problem then is solved using the Matlab implementation of CasADi~\cite{Ande19}. The resulting state response emanating from the optimal control is then simulated on a fine grid of 1001 time grid points, again using the implicit midpoint rule, such that the resulting violation of the energy balance is below three percent.

\subsection{Penalization of the Hamiltonian}

\begin{figure}[htb]
	\centering
	\includegraphics[width=0.51\columnwidth]{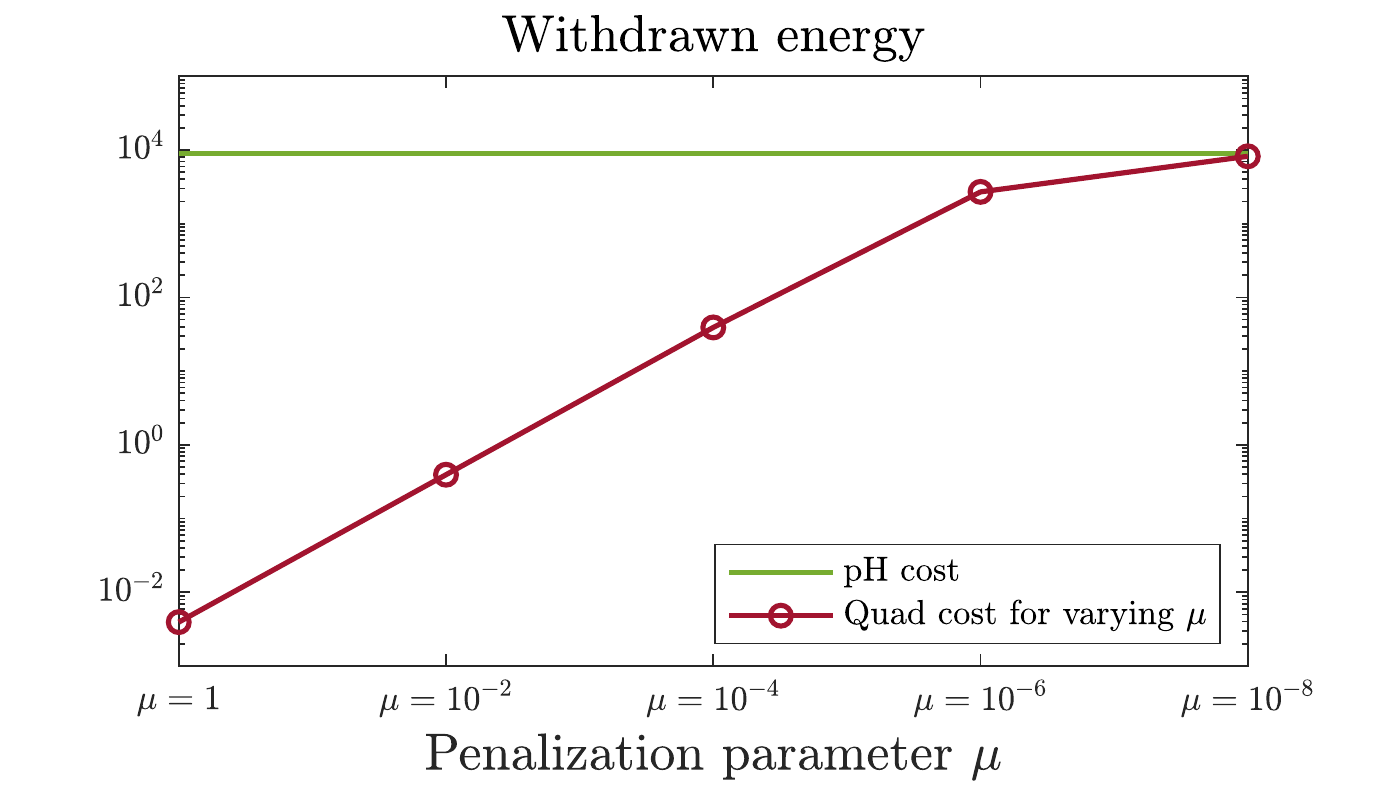} \hspace{-0.7cm}
	\includegraphics[width=0.51\columnwidth]{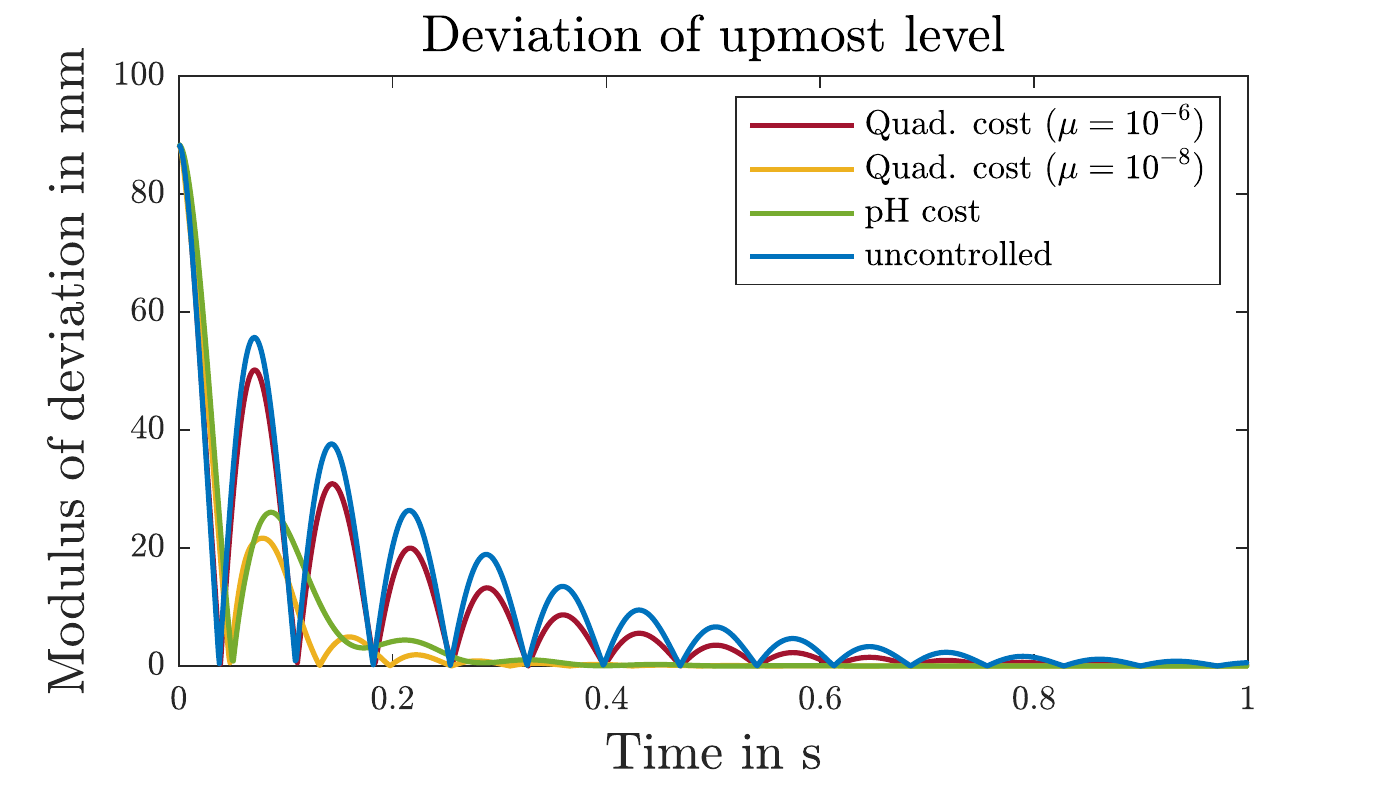}\\
	\includegraphics[width=0.51\columnwidth]{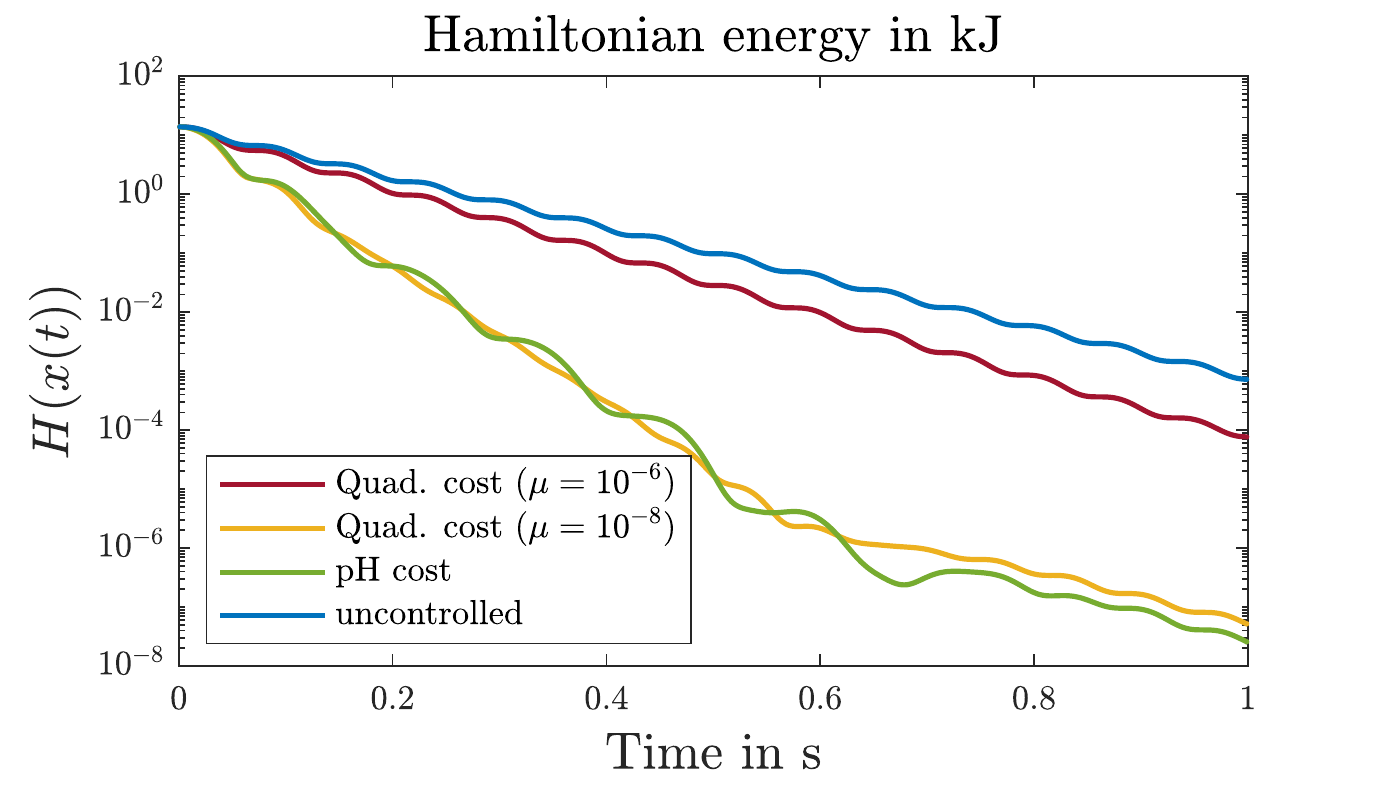} \hspace{-0.7cm}
	\includegraphics[width=0.51\columnwidth]{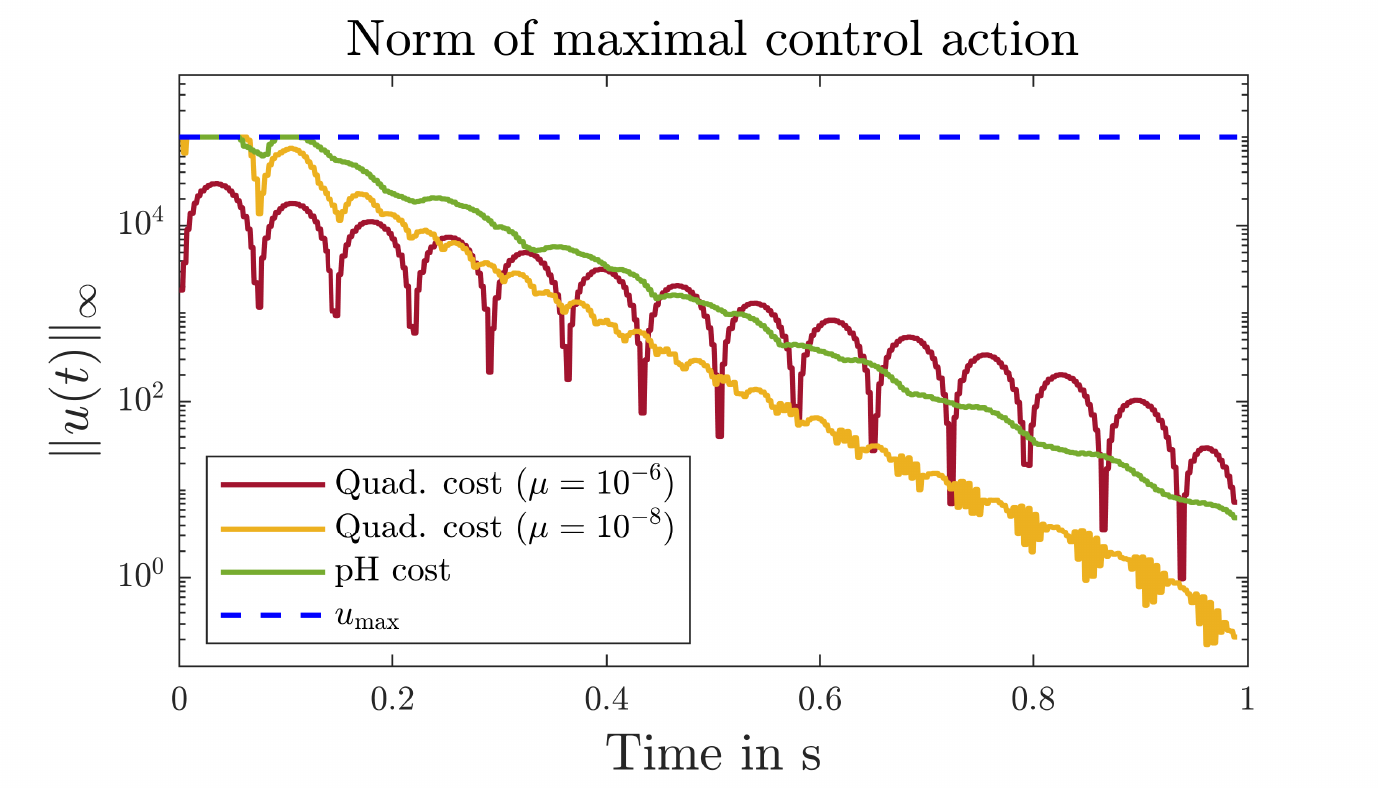}
	\caption{Comparison of the formulations \eqref{eq:lqrocp} and \eqref{eq:phocp} when penalizing the full Hamiltonian.}
	\label{fig:hamiltonian}
\end{figure}

First, we consider penalization of the full Hamiltonian, i.e., we solve OCPs~\eqref{eq:lqrocp} and~\eqref{eq:phocp} with $W=Q$, i.e.,
\begin{align*}
\int_0^T \!x^\top(t) Q x(t) \!+ \!\mu \|u(t)\|^2\,\mathrm{d}t =\! \int_0^T 2H(x(t)) + \mu \|u(t)\|^2\,\mathrm{d}t
\end{align*}
with penalization parameter $\mu > 0$ and 
\begin{align*}
\int_0^T \!x^\top(t) Q x(t) \!+\! u(t)^\top y(t)\,\mathrm{d}t = \!\int_0^T\!2H(x(t)) + u(t)^\top y(t)\,\mathrm{d}t.
\end{align*}
The results in view of vibration damping are shown in Figure~\ref{fig:hamiltonian}.
We observe in the upper left plot that the port-Hamiltonian cost without tuning parameter outperforms all considered penalizations of the quadratic cost in view of the withdrawn energy. The corresponding numerical values are also displayed in Table~\ref{tab:ham_costvals}. Further, in the second and third plot of Figure~\ref{fig:hamiltonian}, it can be seen that the quadratic cost for $\mu = 10^{-8}$ as well as the port-Hamiltonian cost introduce significant additional damping into the system. In contrast, the quadratic cost for $\mu = 10^{-6}$ does not achieve a significant improvement compared to the uncontrolled cost. This reflects again an issue in the quadratic formulation: in order to obtain a good performance in view of vibration damping, a significant amount of tuning is necessary. A reason for this is that the squared control input term lacks physical interpretation. Contrary to this, in the port-Hamiltonian cost functional, both terms are intrinsic to the port-Hamiltonian structure and linked by the energy balance.

\begin{table}[htb]
	\centering
	\resizebox{\columnwidth}{!}{
		\begin{tabular}{|c|c|c|c|c|}\hline
			&Withdrawn energy&Dissipated energy & Remaining energy& Initial energy\\
			&$-\int_0^T u(t)^\top y(t) \,\mathrm{d}t$ &$\int_0^T \|R^{1/2}Qx(t)\|^2\,\mathrm{d}t$ & $H(x(T))$ & $H(x(0))$\\\hline
			uncontrolled&0\,J&13895\,J & $10^{-1}$\,J& 13906\,J\\\hline
			Quad.\ cost \eqref{eq:lqrocp}, $\mu=10^{-6}$&2705\,J&11194\,J& $10^{-2}$\,J& 13906\,J\\
			Quad.\ cost \eqref{eq:lqrocp}, $\mu=10^{-8}$&8221\,J&5712\,J& $10^{-5}$\,J& 13906\,J\\\hline
			pH cost \eqref{eq:phocp}&9019\,J&4882\,J & $10^{-5}$\,J& 13906\,J\\\hline
	\end{tabular}}
	\caption{\centering Penalization of full Hamiltonian: Comparison of supplied and dissipated energy.}
	\label{tab:ham_costvals}
\end{table}
In the lower right plot of Figure~\ref{fig:hamiltonian} we also observe that for the quadratic cost and the choice $\mu = 10^{-6}$ the control constraints are inactive over the full horizon. The reduced penalization $\mu = 10^{-8}$ then leads to active control constraints on the initial part of the horizon, similar to the port-Hamiltonian cost functional. In particular this means that in the case of $\mu =10^{-6}$, a control obtained as a Riccati feedback is feasible, whereas this is no longer the case for $10^{-8}$.

\subsection{Penalization of displacements of the upmost level}
\begin{figure}[htb]
	\centering
	\includegraphics[width=0.51\columnwidth]{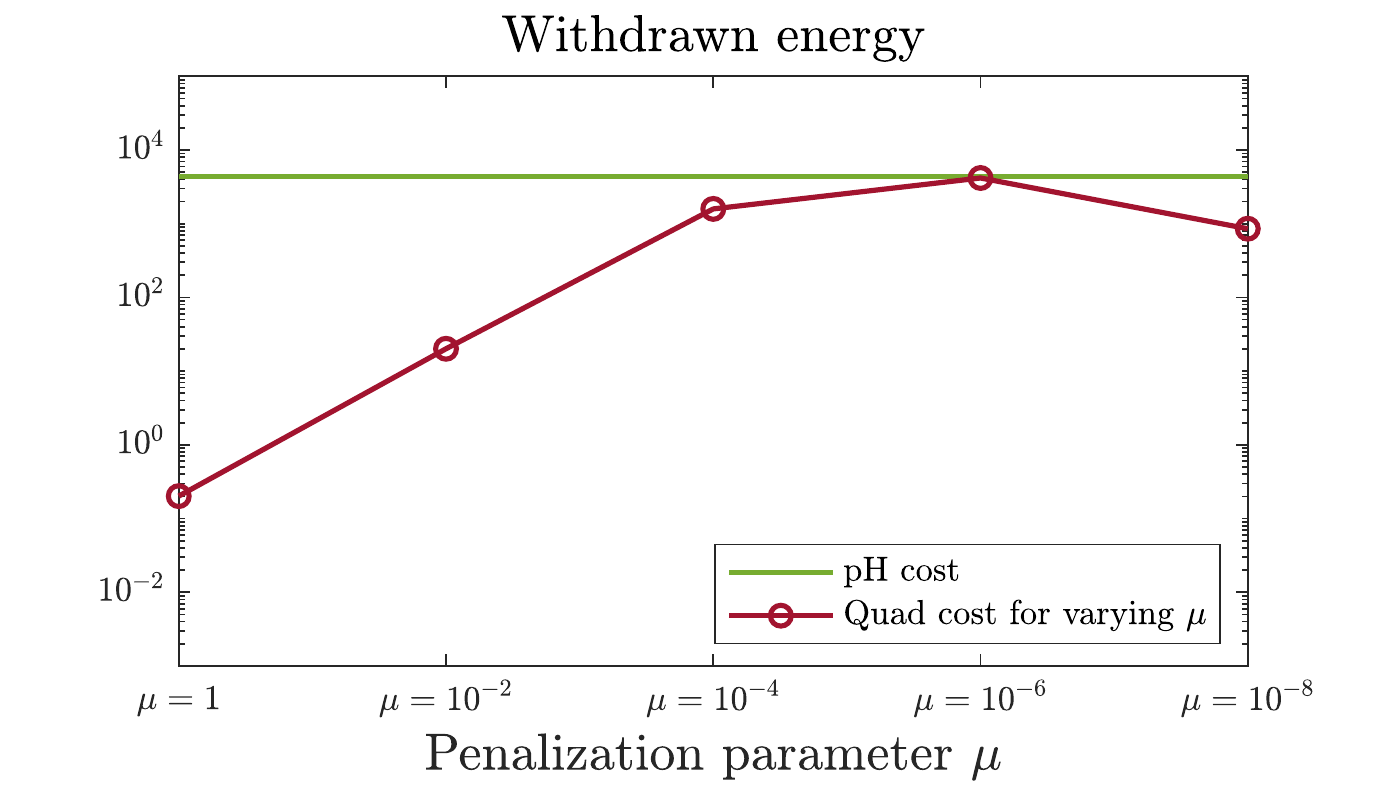}\hspace{-.7cm}
	\includegraphics[width=0.51\columnwidth]{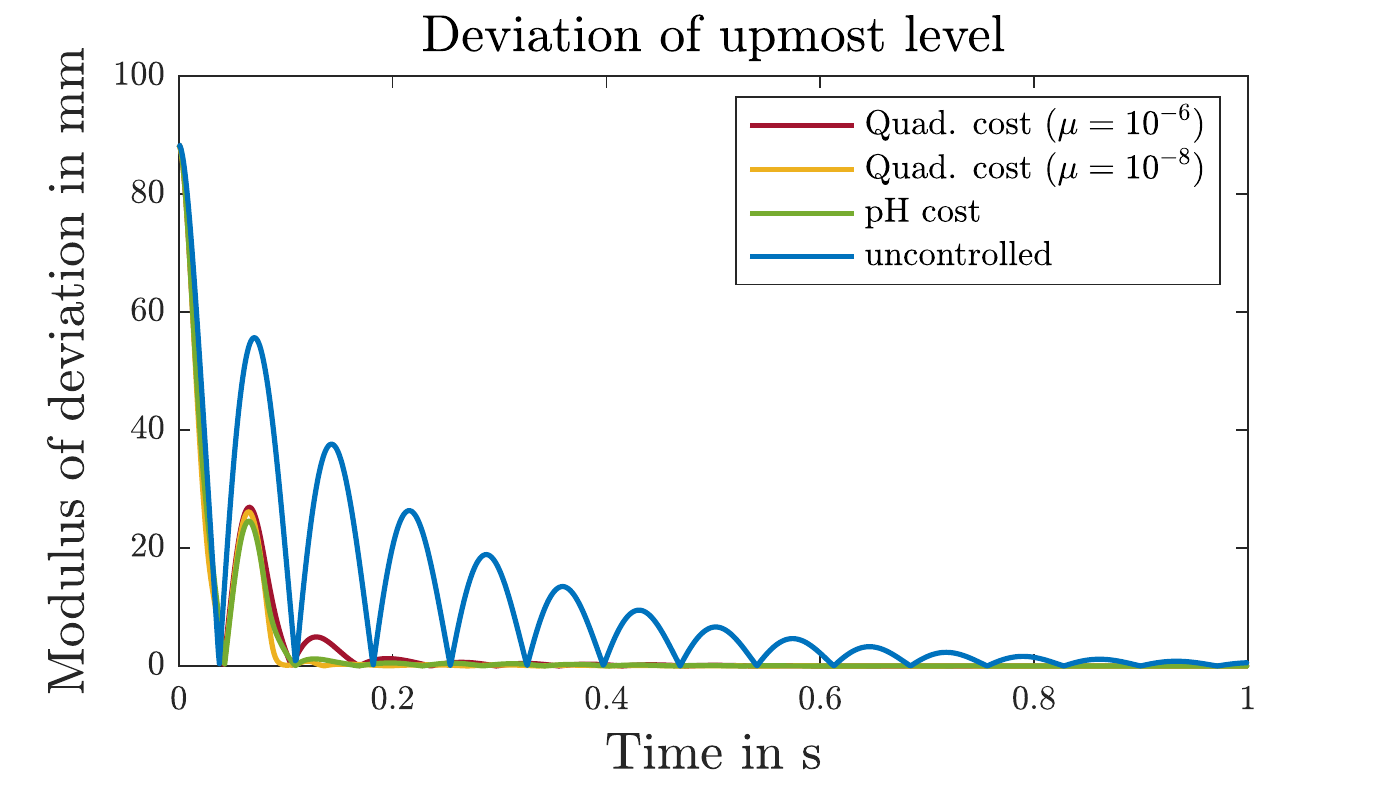}
	\includegraphics[width=0.51\columnwidth]{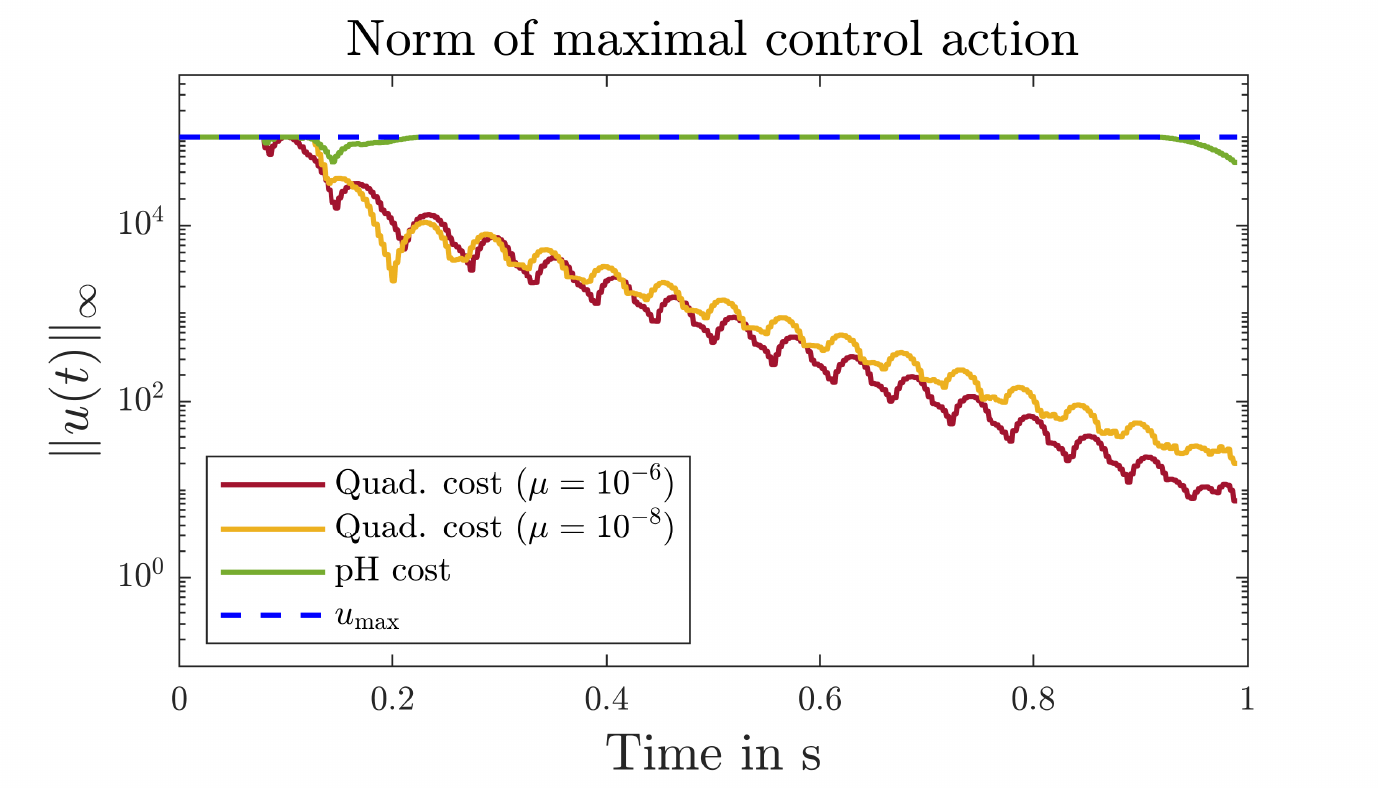}
	\caption{Comparison of the formulations \eqref{eq:lqrocp} and \eqref{eq:phocp} when penalizing the Hamiltonian of the upmost level.}    \label{fig:output_upper_level}
\end{figure}

Here, we consider an example with a weighting matrix for the state which only penalizes the Hamiltonian of the upmost level in the OCPs \eqref{eq:lqrocp} and \eqref{eq:phocp}, i.e., its kinetic and potential energy. The results are depicted in Figure~\ref{fig:output_upper_level}. 
First we see observe again, that the formulation with the port-Hamiltonian cost leads to a significantly higher withdrawn energy. In particular, we observe that for the smallest penalization parameter, the performance of the quadratic cost decreases again. This is contrary to the upper plot in Figure~\ref{fig:hamiltonian}, where we observed a monotone behavior of the withdrawn energy in the penalization parameter. The precise values are depicted in Table~\ref{tab:costvals}. 

\begin{table}[htb]
	\centering
	\resizebox{\columnwidth}{!}{
		\begin{tabular}{|c|c|c|c|c|}\hline
			&Withdrawn energy&Dissipated energy & Remaining energy& Initial energy\\
			&$-\int_0^T u(t)^\top y(t) \,\mathrm{d}t$ &$\int_0^T \|R^{1/2}Qx(t)\|^2\,\mathrm{d}t$ & $H(x(T))$ & $H(x(0))$\\\hline
			uncontrolled&0\,J&13895\,J & $10^{-1}$\,J& 13906\,J\\\hline
			Quad.\ cost \eqref{eq:lqrocp}, $\mu=10^{-6}$&$4174$\,J&9812\,J& $10^{-4}$\,J& 13906\,J\\
			Quad.\ cost \eqref{eq:lqrocp}, $\mu=10^{-8}$&855\,J&13295\,J& $10^{-4}$\,J& 13906\,J\\\hline
			pH cost \eqref{eq:phocp} &4359\,J&9646\,J & $10^{-2}$\,J& 13906\,J\\\hline
	\end{tabular}}
	\caption{\centering Penalization of Hamiltonian corresponding to the upmost level. Comparison of supplied and dissipated energy.}
	\label{tab:costvals}
\end{table}
\newpage
Further, in the middle plot of~Figure~\ref{fig:output_upper_level}, we see that the quadratic cost with the depicted penalization parameters and the port-Hamiltonian cost induce sufficient damping in the system. We note that this is not true for higher damping parameters, where the higher control cost interferes with the goal of damping. We stress again, that such a tuning is not necessary for the port-Hamiltonian cost. 




\section{Conclusion and outlook}\label{sec:conclusion}
We have shown that energy-optimal vibration damping for adaptive high-rise structures can be performed by taking the port-Hamiltonian structure into account in the cost functional. To this end, we proved that optimal states exhibit a turnpike property and, despite singularity of the OCP, optimal controls are characterized by means of the state and corresponding adjoint variable. We provided two numerical examples for different choices of the state cost, showing that the proposed formulation leads to an increased energy efficiency of the controller, while still achieving 
vibration damping.

In future work, the obtained results may be leveraged for (economic) model predictive control in a realistic scenario, e.g., by controlling the high-rise building depicted in Figure~\ref{fig:demonstrator}. To this end, structure-preserving model order reduction techniques~\cite{SchwScha22} have to be employed to obtain a real-time feasible controller.

Whereas Riccati theory could not directly be applied due to the absence of a definite quadratic control penalization in the cost function, the more general framework of Lur'e equations as presented, e.g., in \cite{ReisVoig19} yields a promising alternative for future works to avoid the computation of Drazin or pseudo inverses~\cite{FaulKirch23}.

The developed energy-optimal controller can also be beneficial for energy-harvesting in adaptive structures, where the structure's energy is not simply dissipated in the actuators, but actually is used to build up hydraulic pressure that can be stored and used elsewhere within the same or even a neighbouring structure to drive the actuators. This is expected to increase the energy-efficiency of adaptive structures considerably.

Last, instead of subspace turnpikes as considered in this work, manifold turnpikes~\cite{FaulFlas22} are crucial when going towards more complex models of adaptive structures involving nonlinearities, cf.\ \cite{PhilScha23} for recent results in the context of nonlinear irreversible port-Hamiltonian systems.

\section{Acknowledgements}
The authors would like to thank the referees for the valuable suggestions and their careful reading, which in particular improved the statement of Theorem~\ref{prop:u_xla}. Further, they are grateful to Benedikt Oppeneiger (TU Ilmenau) for providing valuable feedback.

\bibliographystyle{plain}
\bibliography{references.bib}

\end{document}